# Superlens induced loss-insensitive optical force

Xiaohan Cui, Shubo Wang, and C. T. Chan*

*Department of Physics and Institute for Advanced Study, The Hong Kong University of Science and Technology, Hong Kong, China.*

**Abstract**

A slab with relative permittivity $\varepsilon = -1 + i\delta$ and permeability $\mu = -1 + i\delta$ has a critical distance away from the slab where a small particle will either be cloaked or imaged depending on whether it is located inside or outside that critical distance. We find that the optical force acting on a small cylinder under plane wave illumination reaches a maximum value at this critical distance. Contrary to the usual observation that superlens systems should be highly loss-sensitive, this maximum optical force remains a constant when loss is changed within a certain range. For a fixed particle-slab distance, increasing loss can even amplify the optical force acting on the small cylinder, contrary to the usual belief that loss compromises the response of supenlens.

**1. Introduction**

The study of optical force has progressed significantly since Ashkin's pioneering work on the optical manipulations of small particles using focused laser beams [1]. Recently, a lot of interests are focused on the effect of the environment which can significantly enhance and/or modify the optical force acting on an object [2-5]. It is quite surprising that even a simple ordinary slab can dramatically modify the optical force acting on a nearby particle. For example, it is known that an ordinary slab can induce a lateral optical force on a chiral particle [2]. A natural question is what if the slab itself has extraordinary properties. Do we see unique optical forces if a particle is placed near a slab carrying unusual constitutive parameters such as those of a "Veselago slab" with $\varepsilon = \mu = -1$? In the ideal zero absorption limit, such a slab behaves as a "perfect lens" [6-8]. However, absorption should be taken into account in realistic situations, and we call a lens with the constitutive parameters $\varepsilon = \mu = -1 + i\delta$ a "superlens". It is known that even a very small value of $\delta$ can cause significant deviation from the "perfect lens" behavior [9-12]. As such, the absorption cannot be ignored and must be considered explicitly as just about any phenomena associated with the "perfect lens" are very dependent on $\delta$. There are also subtle effects at the $\delta \to 0$ limit. For example, a "Veselago slab" actually behaves as a cloak in the $\delta \to 0$ limit ($\delta$ arbitrarily small but non-zero), so that any small object located within $d/2$ ($d$ is the thickness of the slab) from the lens will be cloaked due to the strong suppression effect of evanescent waves induced on the slab boundary [11]. In fact, a superlens slab generally behaves somewhere between a cloak and a lens in the sense that it will cloak a small particle within $d/2$ region but form an image when the object is located beyond the $d/2$ region [13]. The exact $d/2$ distance marks a special position where the functionality of a superlens changes from cloaking to imaging. At that critical distance of $d/2$, the gradient of the total field acting on the particle diverges in the ideal case (when the particle can be represented as a point dipole). If we examine the standard analytic expression of the optical force acting on a dipolar particle ($F_i = \text{Re}[\alpha E_j \partial_i E_j^*]$, where $\alpha$ is the particle polarizability), a diverging field gradient implies a diverging force but on the other hand, the induced electric dipole moment of the particle also undergoes a dramatic change from almost zero to a finite value at that distance. In a way, this is a zero times infinity type problem and the question then is: what exactly is the optical force?

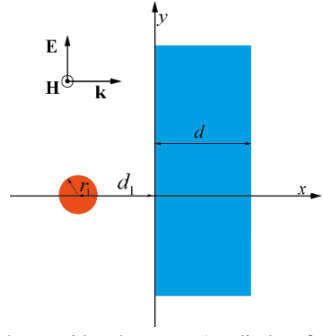

Fig. 1. Schematic picture of the considered system. A cylinder of radius $r_1$ (orange solid circle) is placed at a distance $d_1$ on the left side of a slab (blue) with thickness $d$, permittivity $\varepsilon_s$ and permeability $\mu_s$. The wave vector of the incident plane wave is along $x$ direction.

## 2. Methods and Results

The configuration of the system is illustrated in Fig.1, where a subwavelength cylinder is located on the left-hand side of a superlens slab that is infinitely extended on the $yz$ plane. The radius of the cylinder, the distance from the center of the cylinder to the left surface of the slab and the thickness of the slab are $r_1$, $d_1$ and $d$, respectively. The relative permittivity and permeability of the slab are $\varepsilon_s$ and $\mu_s$, respectively. The system is illuminated by an electromagnetic (EM) plane wave coming from the left whose wave vector is perpendicular to the slab and the magnetic field is polarized along the $z$ direction.

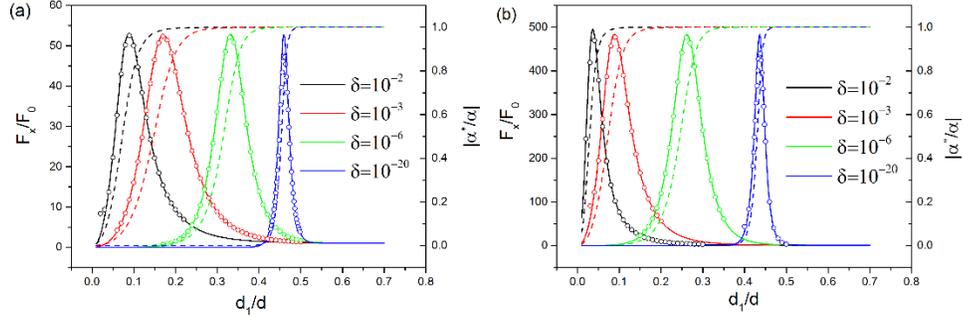

Fig. 2. Normalized optical force $F_x / F_0$ and the effective polarizability $|\alpha^* / \alpha|$ as a function of the particle-slab distance $d_1 / d$. $F_x$ and $F_0$ are the force along $x$ direction with and without the slab. $\alpha$ is the bare polarizability of the particle in the absence of the slab, and $\alpha^*$ is defined in the text. The solid line represents the analytical results obtained using dipole approximation and the circles denote full-wave calculation results using the Maxwell stress tensor approach. The dashed line represents effective polarizability of the cylinder. The slab has thickness $d = \lambda$ and $\varepsilon_s = \mu_s = -1 + i\delta$, where $\lambda = 2000$nm is the wavelength. The cylinder ($r_1 = 0.01\lambda$) is made of metal with $\varepsilon_c = -2, \mu_c = 1$ for (a) and dielectric with $\varepsilon_c = 2, \mu_c = 1$ for (b).

We calculate the optical force acting on the cylinder using a full-wave method and an analytical Green's function method. For the full-wave method [14-15], the force is obtained by integrating the Maxwell stress tensor [$\vec{\mathbf{T}} = \varepsilon_0 \mathbf{EE} + \mu_0 \mathbf{HH} - 1/2(\varepsilon_0 \mathbf{E} \cdot \mathbf{E} + \mu_0 \mathbf{H} \cdot \mathbf{H})\vec{I}$] around a closed path enclosing the cylinder and taking the time average. For the analytical method which will be introduced in the following, the results can be evaluated simply by calculating the field analytically using dipole approximation and then the optical force using $F_i = \text{Re}[\alpha E_j \partial_i E_j^*]$.

We first investigate the effect of absorption on the optical force and the calculated results using full-wave method are shown as open circles in Fig. 2(a) for a metallic cylinder ($\varepsilon_c = -2, \mu_c = 1$) and for a dielectric cylinder ($\varepsilon_c = 2, \mu_c = 1$) in Fig. 2(b). The optical forces here are normalized to the photon pressure $F_0$ in the absence of the slab. We note that the full wave results can be considered as "exact" except for the negligibly small numerical truncation errors. We see that the full-wave results denoted by the circles agree well with the analytical results which are represented by solid lines, meaning that the response of the cylinder is dominated by electric dipole response. The good agreement is quite expected as $r_1 = 0.01\lambda$. As a function of the distance away from the slab, the optical force is essentially zero near the surface ($d_1 \to 0$), reaches a peak value and goes to the limit of $F_x \to F_0$ ($F_0$ is optical force due to photon pressure in the absence of the slab) as $d_1 > d/2$. The sign is positive meaning that the force is in the same direction as the incident photon momentum. The force has a peak value whose position depends on the value of absorption. The peak becomes narrower and its position approaches $d_1/d = 0.5$ as absorption is reduced. In the ideal case of $\delta \to 0$ the optical force vanishes for $d_1/d < 0.5$ due to cloaking effect while its value approaches the ordinary photon pressure $F_0$ for $d_1/d > 0.5$. The peak locates at the point where the dipole changes from being cloaked to being imaged. The most interesting point here is that the peak value of the optical force does not change when the loss is increased from $\delta = 10^{-20}$ to $\delta = 10^{-3}$, which is rather unusual as the properties of a superlens are known to be extremely sensitive to the existence and/or variation of absorption. By comparing Fig. 2(a) (cylinder with $\varepsilon_c < 0$) and Fig. 2(b) (cylinder with $\varepsilon_c > 0$), we see that this absorption-invariant maximum force phenomena exists independent of the property of cylinder while the value of the maximum force (the enhancement) depends on the cylinder property. In addition, this peak value does not depend on the slab thickness $d$ as shown in Fig. 3. Another point worth noting is the strong enhancement of the optical force compared to that in free space.

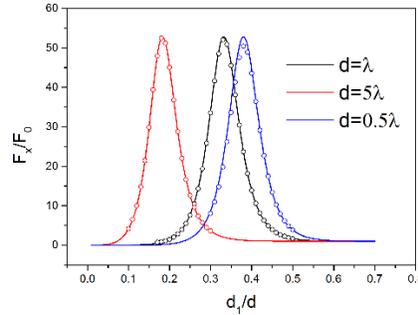

Fig. 3. Optical force acting on a metal cylinder ($\varepsilon_c = -2, \mu_c = 1$) in front of a slab ($\varepsilon_s = \mu_s = -1 + i10^{-6}$) with different thicknesses. The other parameters are the same as in Fig. 2.

In the following, we will derive the analytic formula for the optical force in order to understand why the maximum force is essentially independent of absorption. Under the long wavelength condition, the cylinder can be represented by a passive in-plane electric dipole with the dynamic dipole polarizability $\alpha = 8ia_1\varepsilon_0/k_0^2$ [16], where $k_0$ is the wave vector in vacuum, and $a_1$ is the electric dipole term of Mie scattering coefficient for cylinder [17]. The induced dipole moment is given by

$$P_y = \alpha E_y^{loc} = \alpha\left(E_y^{inc} + E_1^{ref}\right) = \alpha\left(E_y^{inc} + P_y G_{yy}^{ref}\right), \tag{1}$$

where $E_y^{loc}$ and $E_y^{inc}$ are the y component of the local field (including incident electric field and reflected field from the slab due the passive dipole itself ) and the incident field, respectively. In Eq. (1), the reflected field is mainly contributed by the excited dipole because the incident plane wave has negligible reflection in the case of small $\delta$ due to the good impedance matching. $G_{yy}^{ref}$ is the yy component of the reflection part of the dyadic Green's function and it can be expressed as $G_{yy}^{ref} = (-i/4\pi\varepsilon_0)\int_{-\infty}^{\infty} dk_p \exp(2ik_x d_1) k_x R(k_p)$ [14], where $R(k_p)$ is the reflection coefficient of the slab. Here $k_p \equiv k_y$ is the parallel component of the wave vector and $k_x = \sqrt{k_0^2 - k_p^2}$ is the component along the x direction. Eq. (1) can now be rewritten as

$$P_y = \frac{\alpha}{1-\alpha G_{yy}^{ref}} E_y^{inc} = \alpha^* E_y^{inc}, \qquad (2)$$

where $\alpha^*$ is the effective polarizability of the cylinder. The time-averaged optical force along the x direction can be represented as $F_x = 1/2 \text{Re}[P_y^* \partial_x E_y^{loc}]$. Substituting Eq. (2) into the this equation, we obtain

$$F_x = \left|E_y^{inc}\right|^2 \frac{\text{Re}\left\{\alpha^*\left[ik_0\left(1-\alpha G_{yy}^{ref}\right)+\alpha\partial_x G_{yy}^{ref}\right]\right\}}{2\left|1-\alpha G_{yy}^{ref}\right|^2}, \qquad (3)$$

where $\partial_x G_{yy}^{ref}$ is the derivative of $G_{yy}^{ref}$ of x at $-d_1$. This equation is used to produce the analytical results shown as solid lines in Figs. 2-4. The effective polarizabilities [Eq. (2)] are plotted as dashed lines in Fig. 2. If $\alpha^*$ drops to zero, the particle is "cloaked" by slab. In the $\delta \to 0$ limit, $\alpha^*$ is a step function jumping abruptly at . For a finite value of $\delta$, $\alpha^*$ becomes a smooth function and the "cloaking region" shrinks towards the slab surface. We found that the position of the force maximum coincides with this critical distance where the effective polarizability rises from zero to one.

Now we try to find an expression for the maximum force. For the $H_z$ polarization (magnetic field along z direction) and an incident plane wave of the form $H_z \exp(ik_x x + ik_y y)$, the reflection coefficient of the slab can be expressed as [13]

$$R(k_p) = \frac{-(1-\eta^2)\left(e^{ik'_x d} - e^{-ik'_x d}\right)}{(1+\eta)^2 e^{-ik'_x d} - (1-\eta)^2 e^{ik'_x d}}, \qquad (4)$$

where $\eta = \varepsilon_0 k'_x / \varepsilon_s k_x$, and $k'_x = \sqrt{\varepsilon_s \mu_s k_0^2 - k_p^2}$ is the perpendicular component of the wave vector inside the slab. For a slab with $\varepsilon_s = \mu_s = -1 + i\delta$ ($\delta \ll 1$), the reflection coefficient of the evanescent field components ($k_p > k_0$, therefore $k_x = i\sqrt{k_p^2 - k_0^2}$, $k'_x = i\kappa$) can be Taylor-expanded as

$$R(k_p) = \frac{i\delta(-1+e^{2\kappa d})k_p^2}{2(k_p^2 - k_0^2)} - \frac{\delta^2(-1+e^{2\kappa d})k_p^2(k_p^2 - 3k_0^2)}{4(k_p^2 - k_0^2)^2} + O(\delta^3), \qquad (5)$$

where $\kappa = \sqrt{k_p^2 - \varepsilon_s \mu_s k_0^2}$. Eq. (5) indicates that $\text{Im}[R(k_p)]/\text{Re}[R(k_p)] \sim 1/\delta$. We then have $\text{Im}[G_{yy}^{\text{ref}}]/\text{Re}[G_{yy}^{\text{ref}}] \sim \text{Im}[\partial_x G_{yy}^{\text{ref}}]/\text{Re}[\partial_x G_{yy}^{\text{ref}}] \sim 1/\delta$ because evanescent modes are dominant in cloaking region. With $\text{Re}[\alpha]/\text{Im}[\alpha] \sim 1/(k_0 r_1)^2$ and assuming $\delta \ll (k_0 r_1)^2$, we obtain $\alpha G_{yy}^{\text{ref}} = \xi + i\xi'$, $\alpha \partial_x G_{yy}^{\text{ref}} = \zeta + i\zeta'$ where $\xi \approx -\text{Im}[\alpha]\text{Im}[G_{yy}^{\text{ref}}]$, $\xi' \approx \text{Re}[\alpha]\text{Im}[G_{yy}^{\text{ref}}]$, $\zeta \approx -\text{Im}[\alpha]\text{Im}[\partial_x G_{yy}^{\text{ref}}]$ and $\zeta' \approx \text{Re}[\alpha]\text{Im}[\partial_x G_{yy}^{\text{ref}}]$. Therefore, we have $-\text{Im}[\alpha]/\text{Re}[\alpha] \approx \xi/\xi' \approx \zeta/\zeta' \equiv \gamma$. Then Eq. (3) can be rewritten as (for simplicity, we set $|E_y^{\text{inc}}|=1$)

$$F_x = \frac{\text{Re}\{\alpha^*[ik_0(1-\xi-i\xi') + \zeta + i\zeta']\}}{2(1-\xi)^2 + 2\xi'^2} = \frac{\xi' - \gamma + \gamma^2 \xi'}{2(1-2\gamma\xi' + \gamma^2 \xi'^2 + \xi'^2)} \text{Re}(\alpha) k_0. \qquad (6)$$

Eq. (6) has the maximum value at $\xi' = (1+\gamma)/(1+\gamma^2)$:

$$F_x^{\max} = \frac{(1+\gamma^2)}{4}\text{Re}(\alpha)k_0 = \frac{\text{Re}(\alpha)^2 + \text{Im}(\alpha)^2}{4\text{Re}(\alpha)}k_0 = \frac{|\alpha|^2 k_0}{4\text{Re}(\alpha)}. \qquad (7)$$

When normalized to the ordinary photon pressure of $F_0 = 0.5 k_0 \text{Im}[\alpha]$, Eq. (7) becomes

$$\frac{F_x^{\max}}{F_0} = \frac{|\alpha|^2}{2\text{Im}[\alpha]\text{Re}[\alpha]} = \frac{1}{2}\left(\tan\theta + \frac{1}{\tan\theta}\right) \geq 1, \qquad (8)$$

where $\tan\theta = \text{Im}[\alpha]/\text{Re}[\alpha]$. Equation (8) shows that the maximum normalized force only depends on the polarizability of cylinder and is independent of the loss and the thickness of the slab.

The results of Figs. 2-3 show that the maximum optical force can be insensitive to loss, contrary to other phenomena of superlens such as imaging. Such a phenomenon may seem even more intriguing when we consider the force vs. absorption relation for a fixed particle-slab distance, which is shown by the solid black line in Fig. 4. It is seen that the force can be enhanced when the absorption is increased, contrary to our expectation that material loss usually compromises such resonance-related phenomena.

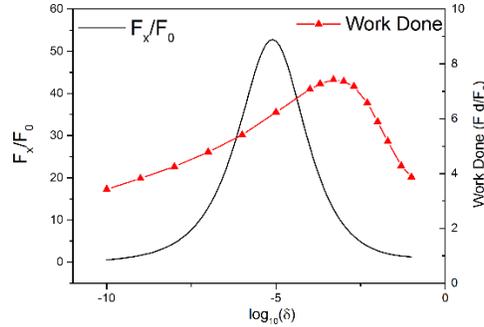

Fig. 4. Normalized optical force (solid black line) acting on a metal cylinder ($\varepsilon_c = -2, \mu_c = 1$) in front of a slab ($\varepsilon_s = \mu_s = -1 + i\delta$) with different values of absorption for fixed particle-slab distance $d_1/d = 0.3$. The red symbol line denotes the work needed to be done in order to move the cylinder from $0.01d$ to $0.5d$ away from the slab. We have normalized it to the work done for ordinary photon pressure without the superlens slab. The other parameters are the same as in Fig. 2.

We also calculate the work needed to be done in order to move the cylinder away from the slab under different loss and the results are shown as the red line in Fig. 4. It can be seen that the work done has a maximum at about $\delta = 5\times 10^{-4}$. While this value of the absorption parameter depends on the details of the system, it is always an "intermediate value" in the sense that it is neither zero or infinity and this coincides with the results shown in Fig. 2, where we see that the integral of $F_x$ over $d_1$ (the area covered by the lines, i.e., the work-done) reaches a maximum when $\delta$ has an intermediate value.

## 3. Conclusion

We studied the optical force acting on a small cylinder located near a superlens slab. We found that the force reaches a maximum value at the critical distance where the slab's behavior is changing from imaging to cloaking. This force maximum is unusual in the sense that it is not sensitive to material loss of the slab and its magnitude remains a constant that is only determined by the cylinder's properties. This is rather unexpected because loss almost always strongly modify the functionalities associated with the perfect lens. The results here are obtained for $H_z$ polarization, but similar phenomenon also exists for $E_z$ polarization (electric field along $z$ direction).

**Acknowledgements**
We thank Prof. Z. Q. Zhang and Dr. M. Xiao for useful discussions. This work was supported by Hong Kong Research Grants Council grant AoE/P-02/12.
*Correspondence to C. T. Chan (phchan@ust.hk)